\documentclass[%
 reprint,
 footinbib,
superscriptaddress,
 amsmath,amssymb,
 aps,
prl]{revtex4-1}

\usepackage{graphicx}
\usepackage{dcolumn}
\usepackage{bm}
\usepackage{braket}
\usepackage{hyperref}
\usepackage[mathlines]{lineno}

\usepackage[usenames]{color}

\renewcommand{\section}[1]{\textit{ #1. } }

\def\lsim{\mathrel{\rlap{\lower4pt\hbox{\hskip1pt$\sim$}}
    \raise1pt\hbox{$<$}}}                

\def\gsim{\mathrel{\rlap{\lower4pt\hbox{\hskip1pt$\sim$}}
    \raise1pt\hbox{$>$}}}                

\begin{document}
 
\title{Bosonic molecules in a lattice: unusual fluid phase from multichannel interactions}

\author{Kevin D. Ewart}
\affiliation{Department of Physics and Astronomy, Rice University, Houston, Texas 77005, USA}
\affiliation{Rice Center for Quantum Materials, Rice University, Houston, Texas 77005, USA}
\author{Michael L. Wall}
\altaffiliation[Current address: ]{The Johns Hopkins University Applied Physics
Laboratory, Laurel, MD 20723, USA}
\affiliation{JILA, NIST and University of Colorado, Boulder, Colorado 80309-0440, USA}
\author{Kaden R. A. Hazzard}
\affiliation{Department of Physics and Astronomy, Rice University, Houston, Texas 77005, USA}
\affiliation{Rice Center for Quantum Materials, Rice University, Houston, Texas 77005, USA}

\date{\today}

\begin{abstract}
We show that multichannel interactions significantly alter the phase 
diagram of ultracold bosonic molecules in an optical lattice. 
Most prominently, an unusual fluid region intervenes between the conventional superfluid and the Mott insulator. In it,
number fluctuations remain but phase coherence is suppressed by a significant factor. This factor can be made arbitrarily large, at least in a two-site configuration.
We calculate the phase diagram using complementary methods, including Gutzwiller mean-field and  
density matrix renormalization group (DMRG) calculations. Although we 
focus on bosonic molecules without dipolar interactions, we expect 
multichannel interactions to remain important for 
dipolar interacting and fermionic molecules. 
\end{abstract}

\maketitle

   
\paragraph{Introduction.}
Over the last several years, multiple species of ultracold nonreactive molecules (NRMs) have been created~\cite{PhysRevA.85.032506,PhysRevA.89.033604,molony:creation_2014,takekoshi:ultracold-RbCs_2014,molony2016production,gregory2016controlling,park:two-photon_2015,park:ultracold_2015,PhysRevLett.116.225306,park2016second,1367-2630-17-3-035003,guo2016creation}, following breakthroughs creating reactive molecules~\cite{Ni_Ospelkaus_08,PhysRevA.86.021602,PhysRevA.89.020702,deiglmayr:formation-LiCs_2008,0953-4075-49-15-152002,moses2017new}. NRMs allow one to avoid the particle loss that occurs in reactive molecules and thereby study rich many-particle physics under conditions where molecular motion is relevant~\cite{carr2009cold,lemeshko2013manipulation,fedorov2016,PhysRevA.88.063632,PhysRevLett.107.115301,PhysRevA.84.033619}, without limiting to regimes protected by the quantum Zeno effect~\cite{zhu_suppressing_2014}, or where motion is completely frozen~\cite{yan_observation_2013,wall:quantum_2015}.
 
However, recent work has shown that the interaction potential between NRMs is more complex than for atoms, with not a few but hundreds of relevant bound states~\cite{mayle:statistical_2012,mayle:scattering_2013,PhysRevA.89.012714,croft2017quantum}. This significantly modifies the Hamiltonian for NRMs in an optical lattice from the usual Hubbard model~\cite{jaksch_bruder_98}, even in the absence of dipolar interactions~\cite{docaj:ultracold_2016,wall:microscopic-derivation-multichannel_2016,wall:beyond_2016}.  While these papers derived the new model, they did not attempt to solve it or characterize its features. Hence, in light of this paper, predictions for the many-body physics of NRMs in optical lattices made in prior work that neglected multi-channel interactions (MCIs) must be revised to account for them, or at least justify neglecting them.

In this paper, we show that MCIs qualitatively alter the  phase diagram of bosonic NRMs in an optical lattice, using the Hamiltonian previously derived in other work.  We use a combination of analytic arguments, Gutzwiller mean-field theory, and density matrix renormalization group (DMRG) calculations to compute and understand the phase diagram of a model that captures the basic features of NRM collisions.

\begin{figure}[h!]
\includegraphics[]{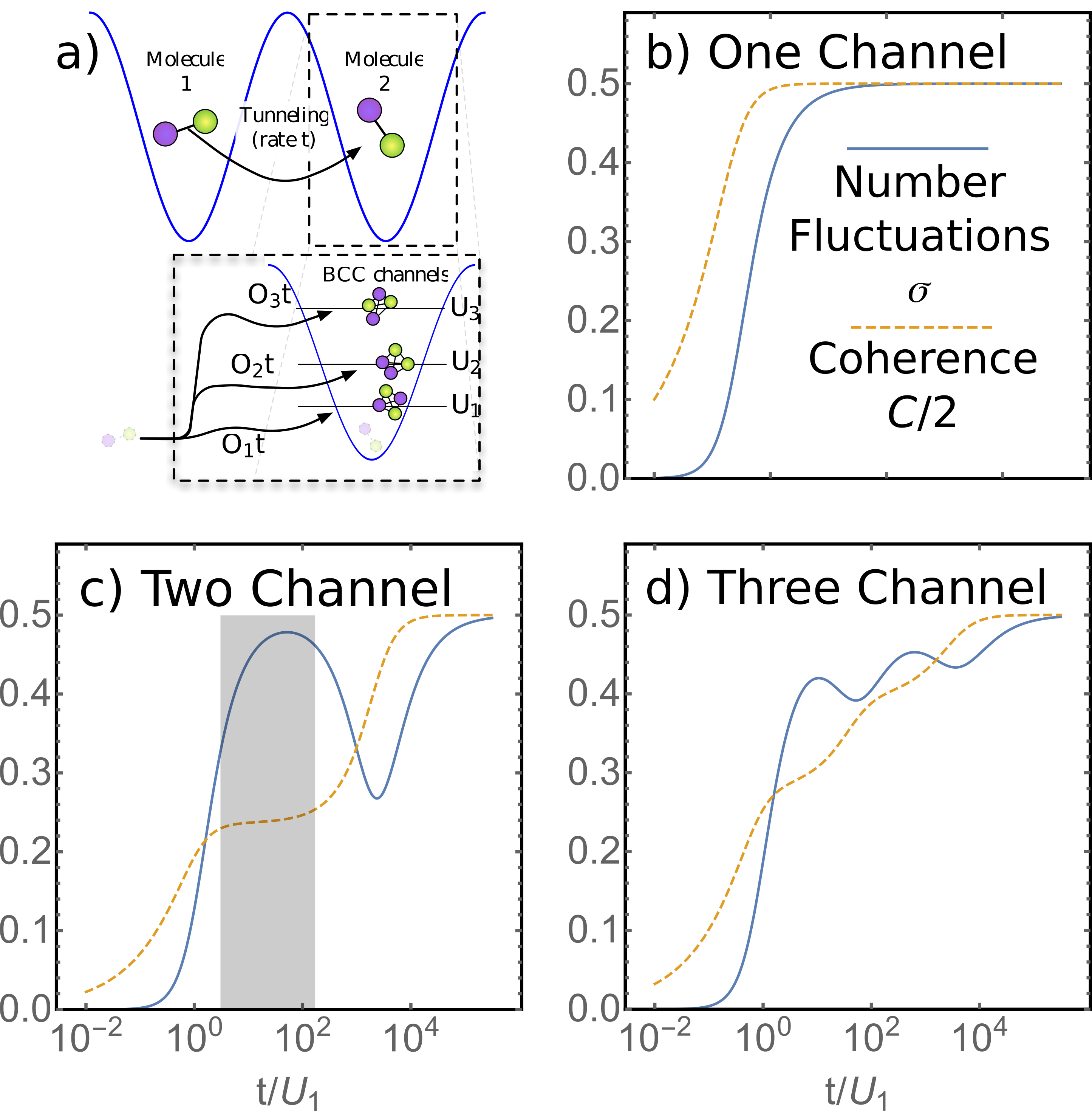}
\caption{(Color online) Nonreactive molecules in a lattice: Hamiltonian and results for two sites. (a) A molecule tunnels into an adjacent singly-occupied site to create one of several two particle states $\ket{2_\alpha}$ with energy $U_\alpha$ where $\alpha$ labels possible channels. The corresponding matrix element is $-tO_\alpha$ where $t$ is the single molecule tunneling rate. (b-d)  Number fluctuations $\sigma$ and halved coherence $C/2$. (b)  Single channel, as for simple atoms. (c) Two channels ($U_2=10^4U_1$, $O_1=0.22, O_2=0.97$) illustrate two major modifications to the phase diagram: (i) A wide intermediate range (shaded) of $t/U_1$ where coherence is significantly suppressed even though number fluctuations are near their large-$t$ values. (ii) Suppressed number fluctuations at the boundary between the intermediate and the large-$t$ regions. (d) Three channel results ($U_2=10^2 U_1$, $U_3=10^4U_1$, $O_1=0.32$, $O_2=0.55$, $O_3=0.77$).}
\label{Twositefig}
\end{figure}

\paragraph{Model.}
Bosonic NRMs in an optical lattice are described by~\cite{docaj:ultracold_2016,wall:microscopic-derivation-multichannel_2016,wall:beyond_2016} 
\begin{equation}
H = -t\sum_{\langle i,j\rangle}c_i^\dagger c_j + \sum_i \sum_\alpha U_\alpha d_{i,\alpha}.
\label{eqnDMRG}
\end{equation}
Here $\sum_{\langle i,j \rangle}$ sums over all nearest neighbors $i$ and $j$, and we define modified creation and annihilation operators $c_i^\dagger \ket{0}_i = \ket{1}_i$, $c_i^\dagger\ket{1}_i = \sqrt{2}\sum_\alpha O_\alpha \ket{2_\alpha}_i$, and $c_i^\dagger\ket{2_\alpha}_i = 0$, where $\ket{n}_i$ is the $n$ molecule ground state on site $i$, $\ket{2_\alpha}_i$ are the two-molecule (doublon) eigenstates on site $i$, and $d_{i,\alpha} = \ket{2_\alpha}_i\bra{2_\alpha}_i$ counts doublons. There can be as many $\sim 1000$ relevant doublon states~\cite{mayle:scattering_2013,docaj:ultracold_2016}.  The $O_\alpha$ and $U_\alpha$ are molecule-dependent parameters whose values were estimated in Refs.~\cite{docaj:ultracold_2016,wall:microscopic-derivation-multichannel_2016,wall:beyond_2016}. The $O_\alpha$'s satisfy $\sum_\alpha O_\alpha^2=1$, and we order the channels $\alpha$ such that $U_1\le U_2\le\dots$.  For simplicity, we neglect dipolar interactions, as is appropriate for zero electric field or for homonuclear molecules, although we expect interesting physics to persist with dipolar interactions. The only assumption for the form of the Hamiltonian that is essential is that the molecular interactions are sufficiently short-ranged compared to the Wannier function length. Under this assumption, an arbitrary multichannel Hamiltonian will have the form we utilized, and we expect this to hold for at least the bialkalis. This is detailed in Refs.~\cite{PhysRevLett.116.135301,wall:microscopic-derivation-multichannel_2016,wall:beyond_2016}.

Figure~\ref{Twositefig}(a) shows a schematic of the processes contained in Eq.~\eqref{eqnDMRG}. A molecule can tunnel from one site to an empty site at rate $t$. Two molecules on two adjacent sites can tunnel onto the same site at a rate $t O_\alpha$. Unlike the usual Hubbard model, there are numerous doublon states, indexed by $\alpha$,  even in the lowest band.

\paragraph{Two Sites.}
 We first consider the problem restricted to two sites with two molecules. Despite its simplicity, this calculation reveals  the essential characteristic features of the MCI model that will appear in the many-site problem.  It is also of direct relevance for understanding small numbers of molecules prepared in tunnel-coupled optical tweezers~\cite{Kaufman_Lester_14}, as has been recently proposed~\cite{liu2017ultracold}.
For this system, the only possible states are $\ket{1,1}=\ket{1}_1\!\ket{1}_2$ and $\ket{\pm_\alpha}=\frac{1}{\sqrt{2}}(\ket{2_\alpha}_1\!\ket{0}_2\pm\ket{0}_1\!\ket{2_\alpha}_2)$. The antisymmetric states $\ket{-_\alpha}$ decouple and are irrelevant for the ground state. The Hamiltonian in the symmetric sector is
\begin{equation}
H = \sum_\alpha  \big[ -2t (O_\alpha\ket{1,1}\!\bra{+_\alpha}+\text{h.c.}) + U_\alpha \ket{+_\alpha}\!\bra{+_\alpha} \big].
\label{eqnTwosite}
\end{equation}
This is an $(N_c+1)\times (N_c +1)$ matrix with $N_c$ the number of channels.  To expose the qualitative phenomena, we choose the $U_\alpha$s to be separated by at least an order of magnitude, and the $O_\alpha$s to be weighted away from the lowest energy $\ket{2_\alpha}$.   

Figure~\ref{Twositefig}(b,c,d) shows  the coherence $C=C_{1,2}$ where $C_{ij}\equiv \sqrt{\langle c_i^\dagger c_j\rangle}$ and number fluctuations $\sigma=\braket{n_i^2}-\braket{n_i}^2$ where $n_i=\ket{1}_i\!\bra{1}_i+2\sum\limits_\alpha\ket{2_\alpha}_i\!\bra{2_\alpha}_i$, 
 in the ground state of Eq.~\eqref{eqnTwosite}  as a function of $t/U_1$ for $N_c=1,2,3$. 
 Some features of the MCI model are similar to the one 
 channel case (that is, the standard Bose-Hubbard model): As $t$ is increased, the number fluctuations 
 and coherence both increase from zero to their maximum around $t\sim U_1$, a two-site precursor of the Mott insulator to superfluid phase transition.

However, two features qualitatively distinguish the MCI model's behavior from the single channel's.
The most significant novel feature is the suppressed coherence in a broad intermediate range of the ``phase diagram" from $t\sim U_1$ to $t\sim U_2$ in the shaded portion of Fig.~\ref{Twositefig}(c).
Another feature is a dip in the number fluctuations appearing around the crossover at $t\sim U_2$. These features are therefore direct consequences of the level structure, as that is the only difference in the models. We will see these survive in the thermodynamic limit.

\begin{figure*}
\includegraphics[]{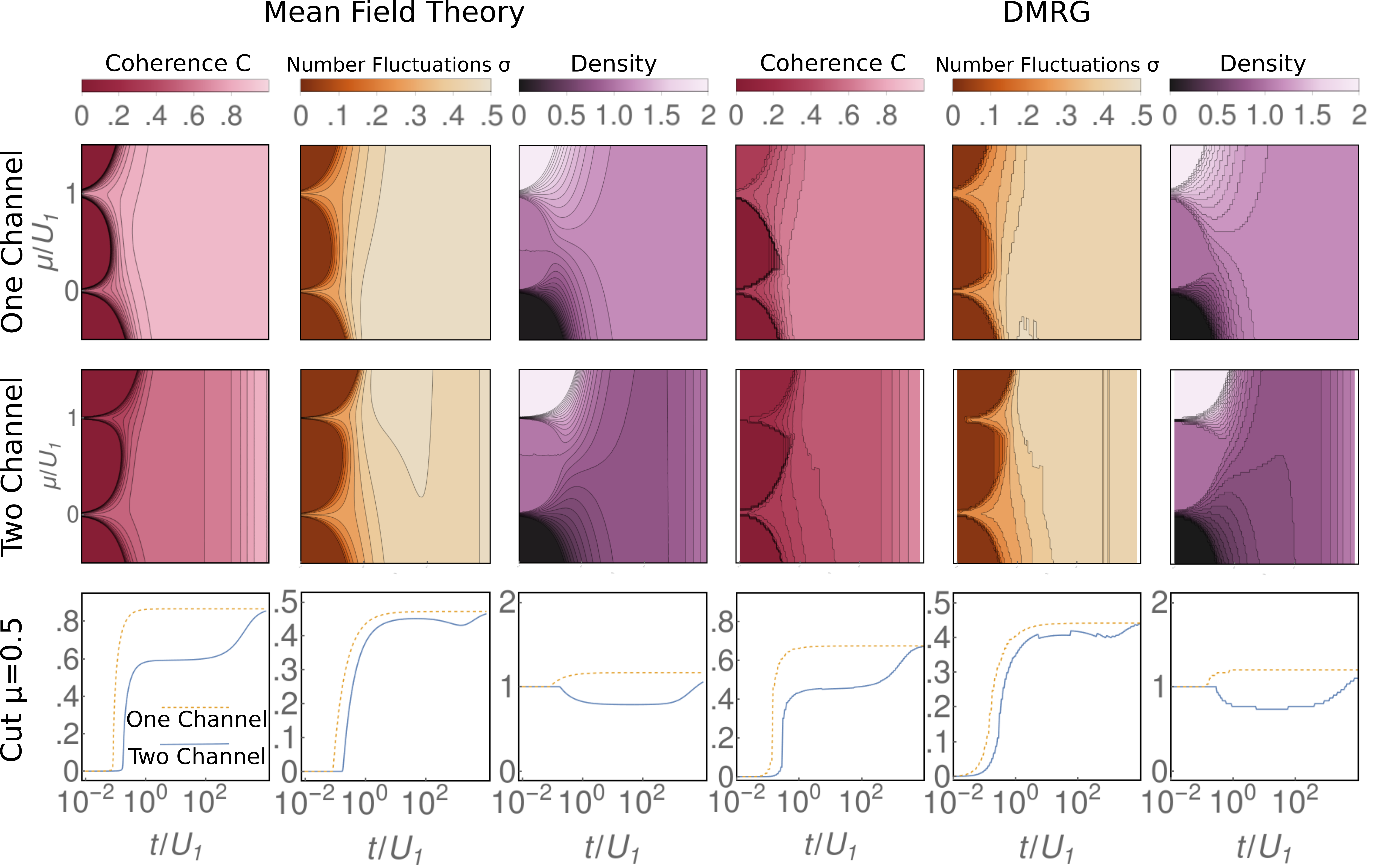}
\caption{(Color online) Coherence $C_{10,20}$, number fluctuations, and density of a 1D chain as a function of $t/U_1$ for Gutzwiller mean field theory (left) and DMRG ($L=$30, right) for one-channel and two-channel, (top to bottom), and cuts at $\mu=0.5$. Parameters are  $U_2=10^4U_1$ and  $O_2=0.89$. Contour lines are spaced in intervals of $0.05$ for $C$ and $\sigma$, and $0.067$ for density. Note that the Mott lobe tilts upwards as $O_1$ is decreased, and the presence of the coherence plateau in the two channel plot. This opacity ratio is at the limit of what our numerical calculations can handle, hence the jagged number fluctuations DMRG cut (see text for details).
\label{fig:MFT-DMRG-phase-diag}}
\end{figure*}

\paragraph{Gutzwiller Mean Field Theory.}
To understand Eq.~\eqref{eqnDMRG}'s ground state beyond the two-site case, we employ two approaches. The first is Gutzwiller mean field theory (MFT), which approximates the ground state as a spatial product state~\cite{fisher1989boson,PhysRevB.44.10328}. Its accuracy for the similar Bose-Hubbard model is discussed in~\cite{PhysRevA.81.013613}. The second is DMRG, which provides accurate results for one-dimensional systems~\cite{schollwock2011density}. We show the qualitative features found with the two-site calculation remain in the more accurate calculations.

The Gutzwiller MFT Hamiltonian is (adding a chemical potential $\mu$ to control the particle number)
\begin{equation}
H = -2t(\eta c + \eta c^\dagger)+\sum_\alpha U_\alpha d_\alpha - \mu n
\label{eqnMFT}
\end{equation}
acting on the single site Hilbert space $\ket{0}$, $\ket{1}$, $\ket{2_\alpha}$.  
Here, $\eta$ is defined to be $C_{i,i+1}=\braket{c}$. For other lattices, including higher dimensions, replace $t$ with $zt/2$, where $z$ is the lattice coordination number. To find the MFT ground state of Eq.~\eqref{eqnMFT}, we iteratively compute $\eta$ until self-consistency with $\langle c\rangle$ in the ground state is achieved~\footnote{Our convergence criterion is $||\eta_m|-|\eta_{m-1}||<p$ for $p=0.001$, where $\eta_m$ is the coherence for the $m$'th step  of the iteration.}
The MFT captures the expected, sharp phase transition between the Mott insulator (small $t$) and superfluid (large $t$), which was smeared out in the two-site case due to its finite size.  The calculated Mott insulator-superfluid phase boundary is similar to the one-channel case. Note that, because of our on-site particle number restriction, only the vacuum, $n=1$, and $n=2$ Mott lobes appear.

Figure~\ref{fig:MFT-DMRG-phase-diag} (left three columns) shows that the key features of the two-site MCI model persist to an infinite system approximated in
the MFT. Namely, the intermediate region $U_1\lsim t \lsim U_2$ displays a  suppressed coherence relative to the single channel model (by $\sim 31\%$), whereas the number fluctuations are barely affected ($\lsim 4\%$). The suppression in coherence can be made even larger by other choices of $O_\alpha$. There is a small dip in number fluctuations at the crossover $t$ and large $\mu$. 

MFT reveals that the MCIs alter the  phase boundary's shape. Because the bandwidth of doublons (hopping on a background of singly occupied sites) decreases as $O_1$ decreases, the Mott lobe tilts up and to the right as $O_1$ decreases. This is similar to the renormalization of doublon tunneling rate that occurs in strongly interacting atoms, as discussed in Refs.~\cite{Duan_05,PhysRevA.81.031602, von_Stecher_Gurarie_11,PhysRevLett.104.090402, PhysRevLett.109.055302,PhysRevA.87.033601}, although there the matrix elements are affected through the band mixing rather than the MCIs.

\paragraph{Numerical solution in one dimension.}
We use DMRG to calculate the density, $C_{L/3,2L/3}$ (chosen to avoid finite size effects), and site-averaged number fluctuations across the phase diagram in a one dimensional (1D) chain~\cite{wall2012out,OSMPS}. This provides highly accurate quantitative results, although some care was required in order to obtain converged results with the MCIs.  We performed calculations for $L=30$ sites and 15 sweeps with discarded weight decreasing from $10^{-3}$ to $10^{-15}$, resulting in maximum bond dimensions of $\sim 350$.  We estimate that these give $C$ and $\sigma$ to within $\sim2$\% in a $O_1 = O_2 = 1/\sqrt{2}$ calculation. For the calculation in the figure, jaggedness is the result of the opacity imbalance being at the limit of what our numerical routines can reasonably handle. However, we are confident in the broad, qualitative features displayed due to its agreement with our MFT calculations within constant factors and accounting for distortion from the Legendre transform in $\mu$ at high $t/U_1$, and due to qualitative similarity to the cited case of known good convergence.

Fig.~\ref{fig:MFT-DMRG-phase-diag} reveals that the features of the MCI Hamiltonian found using the MFT and two-site approximations survive in the DMRG and are thus true properties of Eq.~\eqref{eqnDMRG}, at least in 1D. Specifically, there is an intermediate regime of suppressed coherence, a dip in number fluctuations, and a tilt of the Mott insulator/superfluid phase boundary.

\paragraph{Dependence on $O_\alpha$s and $U_\alpha$s.}
So far, we chose $U_\alpha$s and $O_\alpha$s to illustrate the qualitative effects that emerge from the MCIs. However, the values chosen were not very realistic. Now we examine the dependence on these parameters and incorporate values consistent with expectations~\cite{docaj:ultracold_2016,wall:beyond_2016}.  The consequences of changing $U_\alpha$ are straightforward. We previously  found a series of regions separated at values of tunneling $t\sim U_\alpha$. For example, in the two channel case, $C$ and $\sigma$ increase from zero at small $t$ to a plateau at $t\sim U_1$, and then at $t \sim U_2$ the coherence again increases to the value at which it saturates. Changing the $U_\alpha$ merely changes the locations of these crossovers, and if the $U_\alpha$s are not well-separated, the crossovers blend together. 

\begin{figure}
\includegraphics[]{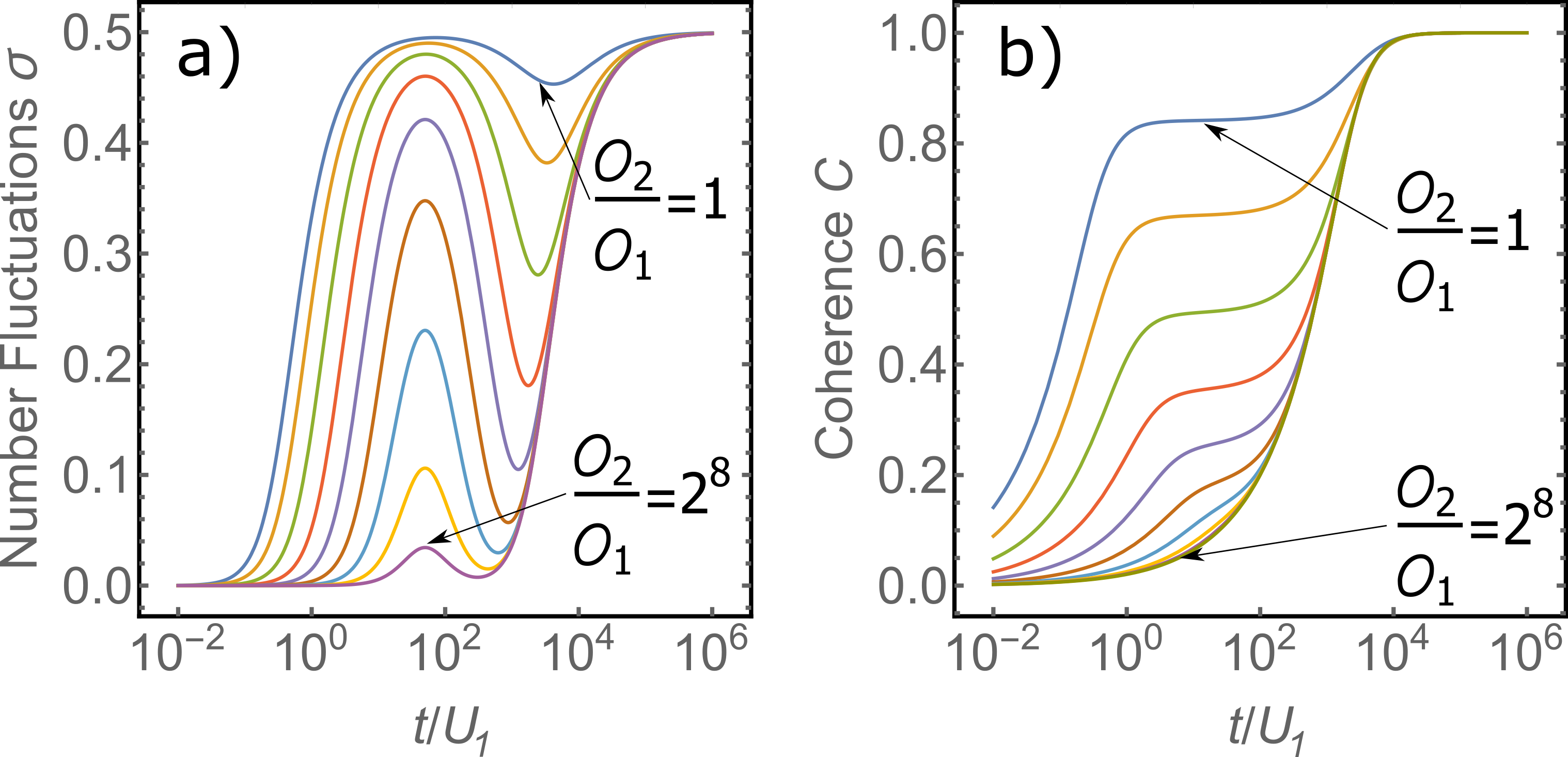}
\caption{(Color online) Effects of changing $O_\alpha$ on (a) number fluctuations and (b) coherence, illustrated by solving two molecules on two sites with two-channel interactions. The ratio $O_2/O_1$ is doubled each step.
\label{fig:extreme-fig}
}
\end{figure}

The consequences of changing $O_\alpha$ are more intricate, as shown in 
Fig.~\ref{fig:extreme-fig}.  Increasing 
$O_2/O_1$ increases 
the $t/U_1$ at which $\sigma$ and $C$ 
initially increase to the intermediate plateau.   The value of $C$ in the intermediate 
plateau decreases with increasing $O_2/O_1$, and $C\rightarrow 0$  as $O_1\rightarrow 0$.   Increasing $O_2/O_1$ also 
causes the ``turn-off" of $\sigma$  to occur at smaller 
$t/U_1$.


These effects can be understood by considering the following limit:  $U_1 \ll U_2$, $O_1 \ll O_2$, and $U_1/O_1 \ll U_2/O_2$. This limit is illuminating because it is contrived to separate the energy scales of the various features, but the features persist beyond this limit. We consider three regimes. (1)  First consider the regime $tO_1 \lsim U_1$. Here $tO_2 \ll U_2$ by our assumptions, and the state $\ket{+_2}$ can be neglected.   Thus, the Hamiltonian reduces to a single channel model with $U=U_1$. 
For $tO_1\ll U_1$, the state is $\ket{1,1}$ -- a Mott insulator with $\sigma=0$ and $C=0$ -- and beyond  $tO_1 \gsim U_1$ it crosses over to $(\ket{1,1}+\ket{+_1})/\sqrt{2}$ with $\sigma=1/2$ and $C=O_1$. Note the $\sigma$ takes its maximum value, but the coherence is suppressed to $O_1$. (2) As $t$ increases further,  $\ket{+_2}$ becomes relevant and it repels the  $\ket{1,1}$ level.   For  $t O_2 \ll U_2$  we can treat the term in Eq.~\eqref{eqnTwosite} that couples $\ket{1,1}$ to $\ket{+_2}$ in second order perturbation theory to find $\Delta E_{1,1} = \frac{-(2tO_2)^2}{U_2}$.
Then the energy gap between $\ket{1,1}$ and $\ket{+_1}$ becomes $U_1+(2tO_2)^2/U_2$. When $(tO_2)^2/U_2$ becomes on the order of $tO_1$, i.e. $tO_2\sim O_1U_2/O_2$, the transition between $\ket{1,1}$ and $\ket{+_1}$ will be sufficiently off-resonant that the ground state is again simply $\ket{1,1}$. Therefore, $C$ and $\sigma$ return to zero. 
(3) Finally, for very large $t$ such that $t O_2 \gsim U_2$,  the coupling of $\ket{+_2}$ to $\ket{1,1}$ dominates any coupling to the $\ket{+_1}$  state, and  we can work with the Hamiltonian projected to $\ket{1,1}$ and $\ket{+_2}$. Consequently, increasing $t$ beyond $t O_2 \sim U_2$, the ground state becomes $(\ket{1,1}+\ket{+,2})/\sqrt{2}$ and thus $\sigma \rightarrow 1/2$ and $C\rightarrow 1$. 

Now that we understand the effects of general $O_\alpha$ and $U_\alpha$, we consider typical values for them. These depend on as yet unknown (either theoretically or experimentally) species-specific molecular properties, as well as other parameters of the experiment. Hence, we will take a statistical approach, in which we sample parameters from an appropriate probability distribution informed by a combination of random matrix theory, transition state theory, and quantum defect theory~\cite{docaj:ultracold_2016,wall:microscopic-derivation-multichannel_2016,wall:beyond_2016}.  As a distribution that roughly captures the features of this approach, we take the energies from a uniform distribution from between $U_\alpha=0$ and $U_\alpha=1000$, and the $O_\alpha$ are sampled from a normal distribution with zero mean then normalized so that $\sum_\alpha O_\alpha^2=1$.
 The most artificial aspect of this choice is that we consider only positive energy states,  although both negative and positive states are equally likely.  This choice is mainly a convenience to preserve stability at zero temperature, but it may be qualitatively appropriate in some regimes and regardless is a first step towards a fuller understanding.

\begin{figure}
\includegraphics[]{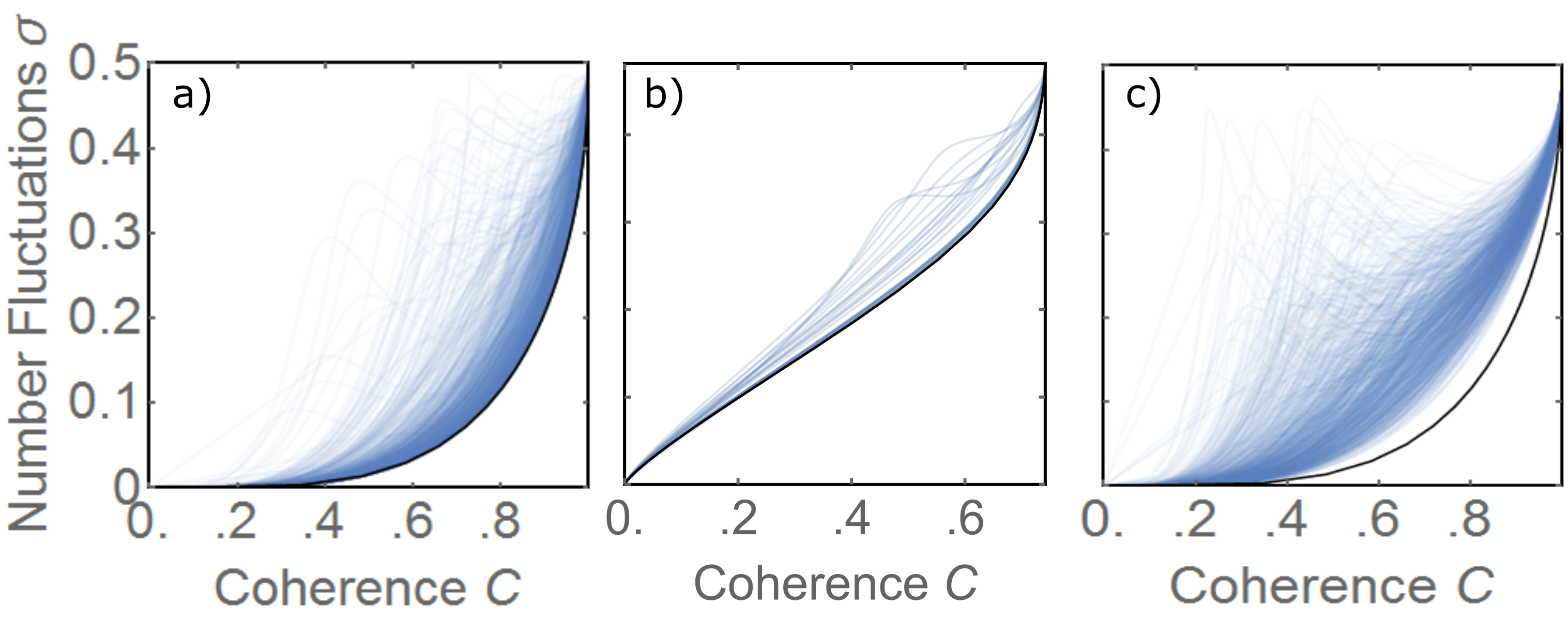}
\caption{(Color online) Effect of random distribution of $O_\alpha$s and $U_\alpha$s. (a) Parametric plots of coherence versus number fluctuations for two molecule, two site, five-channel model. We plot 1000 curves, each from parameters sampled from a physically realistic distribution described in the text.  (b) Same as  (a) for $L=15$ chain with $\braket{n}=1$, calculated using DMRG. (c) Same as (a), but with 100 channels. In all cases, there are strong deviations from the single-channel result (bottom solid line).}
\label{fig:parametric-fig}
\end{figure}

 Figure~\ref{fig:parametric-fig} shows the effects of including realistic $U_\alpha$s and $O_\alpha$s. Rather than plotting the phase diagram, we parametrically plot $C$ versus $\sigma$ for varying $t$ at fixed $n=1$. We plot this way because the phase diagram -- for example the Mott insulator-superfluid boundary -- can fluctuate wildly due to trivial rescalings of the $U_\alpha$. By plotting parametrically we avoid the trivial rescalings: for example, in the single channel case this plot would be independent of $U$.

  Figure~\ref{fig:parametric-fig} reveals that the MCIs have a strong effect on the phase diagram with remnants of the qualitative features that we have already identified. First consider the two-site, $N_c=5$, results shown in Fig.~\ref{fig:parametric-fig}(a). The bottom-most curve is the single channel result, with the Mott insulator in the bottom left and the superfluid in the upper right. All of the MCI curves lie above and to the left of this, demonstrating the suppression of coherence at fixed $\sigma$ in the intermediate phase, just as in the earlier examples. For a typical sample, $\sigma$ can be enhanced by several tens of percent, and up to a factor of 4 in $\sim 10$\% of our samples.  The same behavior is seen in the DMRG calculations [Fig.~\ref{fig:parametric-fig}(b)] with $L=15$.  Finite-size effects suppress the sharp feature in $\sigma$ near $C\sim 0$, which is recovered for $C_{ij}$, $j\gg i$, and $L\to \infty$.  Finally, Fig.~\ref{fig:parametric-fig}(c) shows the two-site case for physically relevant $N_c$, where the bulk of the curves pull away from the $N_c=1$ curve. We expect this to persist in the many-body case, but it is not technically feasible to perform accurate DMRG calculations with this many channels.

\paragraph{Conclusions.} 
We demonstrated that multichannel interactions lead to  characteristic features in the phase diagram of NRMs in an optical lattice, focusing on the simplest case of bosonic molecules without an applied electric field (including homonuclear ground state molecules~\cite{herbig2003preparation,danzl:quantum_2008,danzl:ultracold_2010,
reinaudi:optical_2012,stellmer:creation_2012}). The most striking difference with the usual one-channel Hubbard phase diagram is a large regime at intermediate $t/U$ between the usual Mott insulator and superfluid where coherence is suppressed while number fluctuations remain large. In this sense, the system acts as a normal fluid: particles move, but there is arbitrarily small off-diagonal order.

 This work opens many questions. One major question is whether the steady state of Eq.~\eqref{eqnDMRG} from any realistic initial condition can even support a superfluid.  Another arises from our  omission  of channels with $U_\alpha<0$. The approximate probability distribution of $U_\alpha$, as given in Refs.~\cite{docaj:ultracold_2016,wall:microscopic-derivation-multichannel_2016,wall:beyond_2016}, has equal likelihood of positive and negative $U_\alpha$.  Negative values of $U_\alpha$ will  cause the bosons to trivially clump together on a single site in the ground state in the thermodynamic limit~\cite{oelkers2007ground}, although energy or kinetics suppressing triple occupancy can prevent this, as pointed out~\cite{diehl2010quantum,diehl2010quantumb} for the two-channel case~\cite{PhysRevLett.92.160402,PhysRevLett.93.020405}.  Nevertheless, our calculations are relevant. First, experimental situations are possible where all the $U_\alpha$ are positive, especially when only a  few channels are present. Second, at finite temperature, the system may be prevented from collapse, and the ground state phase diagram may point to features of this  finite temperature phase diagram.  In the future, it will be crucial to understand the phase diagram and dynamics of other systems with multichannel collisions, including  fermionic, spinful, or dipolar molecules~\cite{0034-4885-72-12-126401,doi:10.1021/cr2003568}.

\acknowledgements

We acknowledge Rick Mukherjee and Ian White for conversations.
K.R.A.H. thanks  the  Aspen  Center  for  Physics,  which  is
supported by National Science Foundation grant PHY-1066293,  for its hospitality while part of this work was
performed.   This  work  was  supported  with  funds  from
the  Welch  foundation,  Grant  No.   C-1872, and in part by the Data Analysis and Visualization Cyberinfrastructure funded by NSF under grant OCI-0959097 and Rice University.  K.D.E. acknowledges support by a scholarship from the Physics and Astronomy Department at Rice University.

\bibliography{NRM-PDRefs}

\end{document}